\documentclass[RNAAS]{aastex63}

\shorttitle{IRFM temperatures with GAIA DR2}
\shortauthors{Mucciarelli \& Bellazzini}

\def\bprp{$(\rm BP-RP)_{0}$}
\def\bpg{$(\rm BP-G)_{0}$}
\def\grp{$(\rm G-RP)_{0}$}
\def\bpk{$(\rm BP-K)_{0}$}
\def\rpk{$(\rm RP-K)_{0}$}
\def\gk{$(\rm G-K)_{0}$}

\def\teff{${\rm T_{eff}}$}

\begin{document}

\title{Gaia DR2 colour-temperature relations 
based on infrared flux method results}

\author{Alessio Mucciarelli} \affiliation{Universit\`a di Bologna,
  Dipartimento di Fisica e Astronomia, Via Gobetti 93/2 I-40129
  Bologna, Italy}
\affil{INAF -- Osservatorio di Astrofisica e Scienza dello Spazio di Bologna, Via Gobetti 93/3 I-40129 Bologna,
  Italy}

\author{Michele Bellazzini}
\affil{INAF -- Osservatorio di Astrofisica e Scienza dello Spazio di Bologna, Via Gobetti 93/3 I-40129 Bologna,
  Italy}


\section{}
The ESA/Gaia mission \citep{prusti16} is providing accurate and precise all-sky photometry in three photometric filters named
G, BP and RP \citep{evans18}. The superb quality of this homogeneous photometric database allows a significant improvement in the determination of stellar parameters.
Amongst them, the effective temperature 
( \teff\ ) is the most crucial parameter in the chemical abundance derivation because 
it affects any kind of atomic or molecular line.
Photometric \teff\ are preferred to those derived from excitation and ionisation balances in stellar spectra, in particular at low metallicity 
\citep[see e.g.][]{mb20},  being less affected by  the inadequacies of  the models of stellar atmospheres.

One of the most widely adopted methods to derive  accurate stellar \teff\ is the infrared flux method (IRFM) that  uses the monochromatic flux in the near infrared 
bands. \citet{ghb09} derived \teff\ for a sample of dwarf and giant stars through IRFM and using the 2MASS photometry and provided relations between \teff\ and different broad-band colours based on Johnson-Cousin and 2MASS magnitudes.

In order to extend these relations to colours including the Gaia magnitudes, 
we combined \teff\ derived by \citet{ghb09} for their sample of dwarf and giant stars, 
with the photometry available from the second release of the ESA/Gaia 
mission \citep{brown18,evans18} and the K-band magnitudes from the 2MASS database \citep{2mass}. 
We consider the broad-band colours 
 \bprp\, \bpg\, \grp\, \gk\, \bpk\ and \rpk\ .
We adopted the same colour excess values, E(B-V), provided by \citet{ghb09}.  
The reddening corrections have been performed following  the procedure described in \citet{bab18}, for the Gaia 
filters, and adopting the extinction coefficient by \citet{mccall04} for the K filter.

To derive our color-T$_{\rm eff}$ relations we used only stars with uncertainties $<0.1$~mag in the Gaia magnitudes, but
most of the stars in our sample have uncertainties in the individual magnitudes of about 0.005 mag or less.
We performed a fit for each colour 
(considering separately dwarf and giant stars), using the fitting formula usually adopted in other studies based on IRFM

\begin{equation}
\theta = {\rm b_0}+{\rm b_1}{\rm C}+{\rm b_2}{\rm C^2}+{\rm b_3}{\rm [Fe/H]}+{\rm b_4}{\rm [Fe/H]^2}+{\rm b_5}{\rm [Fe/H]}{\rm C}
\end{equation}

where $\theta$=5040/\teff\ , {\rm C} is the used colour 
and $b_0$,...,$b_5$ are the coefficients of the fit. 
We adopted an iterative 2.5$\sigma$-clipping procedure to remove  outliers. 

Table 1 lists for each colour the colour range, the 
final number of stars used for the fit, the 1$\sigma$ dispersion of 
the fit residuals and the coefficients $b_0$,...,$b_5$.
For the dwarf stars, the residuals have a standard deviation
of about 50-70 K, while it is 50-100 K for the giant stars. 
These dispersions are comparable with those obtained 
for Johnson-Cousin-2MASS colours by \citet{ghb09}.
We note that the colours including the K-band magnitude
provide colour-\teff\ relations with lower 
dispersion of the residuals than
pure Gaia colours, especially for giant stars, probably due to their large baseline in wavelength.

Figures 1-6 show the behaviour of \teff\ by \citet{ghb09} as a function of the adopted colours, 
grouping the stars in four metallicity bins: [Fe/H]$<$--2.5 dex 
(blue points),--2.5$<$[Fe/H]$,$--1.5 (green points), --1.5$<$[Fe/H]$,$--0.5 (red points), 
[Fe/H]$>$--0.5 dex (black points). The polynomial fits calculated with four metallicity values are superimposed
with the same colour-code used for the stars. 

For dwarf stars, the dispersion is similar in all the metallicity bins. 
On the other hand, for the giant stars 
the dispersion is dominated by the stars with [Fe/H]$>$--0.5 that have 
larger uncertainties in the Gaia magnitudes.

We estimated the typical uncertainties in the derived \teff\ due to the uncertainties in the adopted photometric colours. We considered
the average photometric errors in \bprp\ as quoted 
by \citet{evans18}. The behaviour of the \teff\ variations as a function of the G-band magnitude for dwarf and 
giant stars is shown in Fig.~\ref{errteff}. 
For dwarf stars, \teff\ with uncertainties of about 50 and 100 K can be obtained for stars with G$\sim$17 and 18, respectively. For giant stars, the same errors in \teff\ can be obtained at G$\sim$18 and $\sim$18.7. 
For stars fainter than G$\sim$18 and $\sim$18.7 (for dwarf and giant stars, respectively), the uncertainties in \teff\ increase significantly.
Note that the metallicity has a small impact on the derived errors.

These relations allow the exploitation of the all-sky
Gaia DR2 photometry to obtain precise \teff\ measurements, consistent with those derived
from other photometric data via the method described in \citet{ghb09}.

\begin{table}[htbp]
\caption{Coefficients ${\rm b_{0}}$,...,${\rm b_{5}}$ of the colour-\teff relations based on GAIA DR2 magnitudes, together with corresponding colour range, the dispersion of the fit residuals and the number of used stars.}             
\label{tab1}      
\centering                          
\begin{tabular}{c c c c   c c c  c c c}        
\hline\hline                 
{\rm Colour} & {\rm Colour range} & $\sigma_{\rm T_{\rm eff}}$  & {\rm N} & 
    ${\rm b_0}$ & ${\rm b_1}$ & ${\rm b_2}$ & ${\rm b_3}$ & 
    ${\rm b_4}$ & ${\rm b_5}$  \\ 
\hline
	  &  (mag)  &   (K)   &   &    &   &      &     &      &    \\
\hline
	  &    &      &   &  Dwarf stars  &   &      &     &      &       \\
\hline

\bprp\  &    [0.38--1.51]  &  61  &   445  &   0.4988   &  0.4925  &  --0.0287  &   0.0193  &  --0.0017  &  --0.0384  \\
\bpg\   &    [0.17--0.72]  &  77  &   429  &   0.4800   &  1.3160  &  --0.4957  & --0.0086  &  --0.0020  &  --0.0444  \\
\grp\   &    [0.17--0.79]  &  68  &   438  &   0.5623   &  0.5422  &    0.3069  &   0.0367  &  --0.0019  &  --0.0829  \\
\bpk\   &    [0.64--3.24]  &  47  &   454  &   0.5375   &  0.1967  &  --0.0002  &   0.0268  &    0.0006  &  --0.0150  \\
\rpk\   &    [0.34--1.75]  &  54  &   444  &   0.5451   &  0.3739  &  --0.0120  &   0.0289  &    0.0026  &  --0.0185  \\
\gk\    &    [0.52--2.53]  &  51  &   446  &   0.5576   &  0.2191  &    0.0095  &   0.0334  &    0.0014  &  --0.0182  \\

\hline
	  &    &      &   &  Giant stars  &   &      &     &      &       \\
\hline
\bprp\  &    [0.34--1.80]  &  83  &   229  &   0.5403  &   0.4318  &  --0.0085  & --0.0217  &  --0.0032  &    0.0040  \\
\bpg\   &    [0.13--1.00]  & 106  &   218  &   0.5156  &   1.3488  &  --0.6976  & --0.0105  &  --0.0020  &  --0.0181  \\
\grp\   &    [0.21--0.84]  &  86  &   190  &   0.5056  &   0.8788  &    0.0107  &   0.0216  &    0.0023  &  --0.0030  \\
\bpk\   &    [0.69--3.98]  &  52  &   233  &   0.5670  &   0.1829  &  --0.0004  &   0.0030  &  --0.0009  &  --0.0034  \\
\rpk\   &    [0.35--2.26]  &  64  &   235  &   0.5764  &   0.3601  &  --0.0237  &   0.0350  &    0.0000  &  --0.0245  \\
\gk\    &    [0.56--3.06]  &  66  &   230  &   0.5444  &   0.2747  &  --0.0118  &   0.0387  &    0.0024  &  --0.0117  \\

\hline                  
\hline                                   
\end{tabular}
\end{table}

\bibliography{sample63}{}

\begin{thebibliography}{}

\bibitem[Gaia Collaboration, Prusti et al.(2016)]{prusti16} 
Gaia Collaboration, Prusti, T., de Bruijne, J.~H.~J., et al.\ 2016, \aap, 595, A1

\bibitem[Gaia Collaboration, Babusiaux et al.(2018)]{bab18} 
Gaia Collaboration, Babusiaux, C., van Leeuwen, F., et al.\ 2018, \aap, 616, A10

\bibitem[Gaia Collaboration, Brown et al.(2018)]{brown18} 
Gaia Collaboration, Evans, D.W., Rielllo, M., et al.\ 2018, \aap, 616, A4

\bibitem[Gaia Collaboration, Evans et al.(2018)]{evans18} 
Gaia Collaboration, Brown, A.~G.~A., Vallenari, A., et al.\ 2018, \aap, 616, A1

\bibitem[Gonz{\'a}lez Hern{\'a}ndez \& Bonifacio(2009)]{ghb09} 
Gonz{\'a}lez Hern{\'a}ndez, J.~I., \& Bonifacio, P.\ 2009, \aap, 497, 497 

\bibitem[McCall(2004)]{mccall04} 
McCall, M.~L.\ 2004, \aj, 128, 2144

\bibitem[Mucciarelli \& Bonifacio(2020)]{mb20} 
Mucciarelli, A., \& Bonifacio, P.\ 2020, arXiv e-prints, arXiv:2003.07390


\bibitem[Skrutskie et al.(2006)]{2mass} 
Skrutskie, M.~F., Cutri, R.~M., Stiening, R., et al.\ 2006, \aj, 131, 1163 


\end{thebibliography}
\bibliographystyle{aasjournal}

\begin{figure}
\includegraphics[width=\columnwidth]{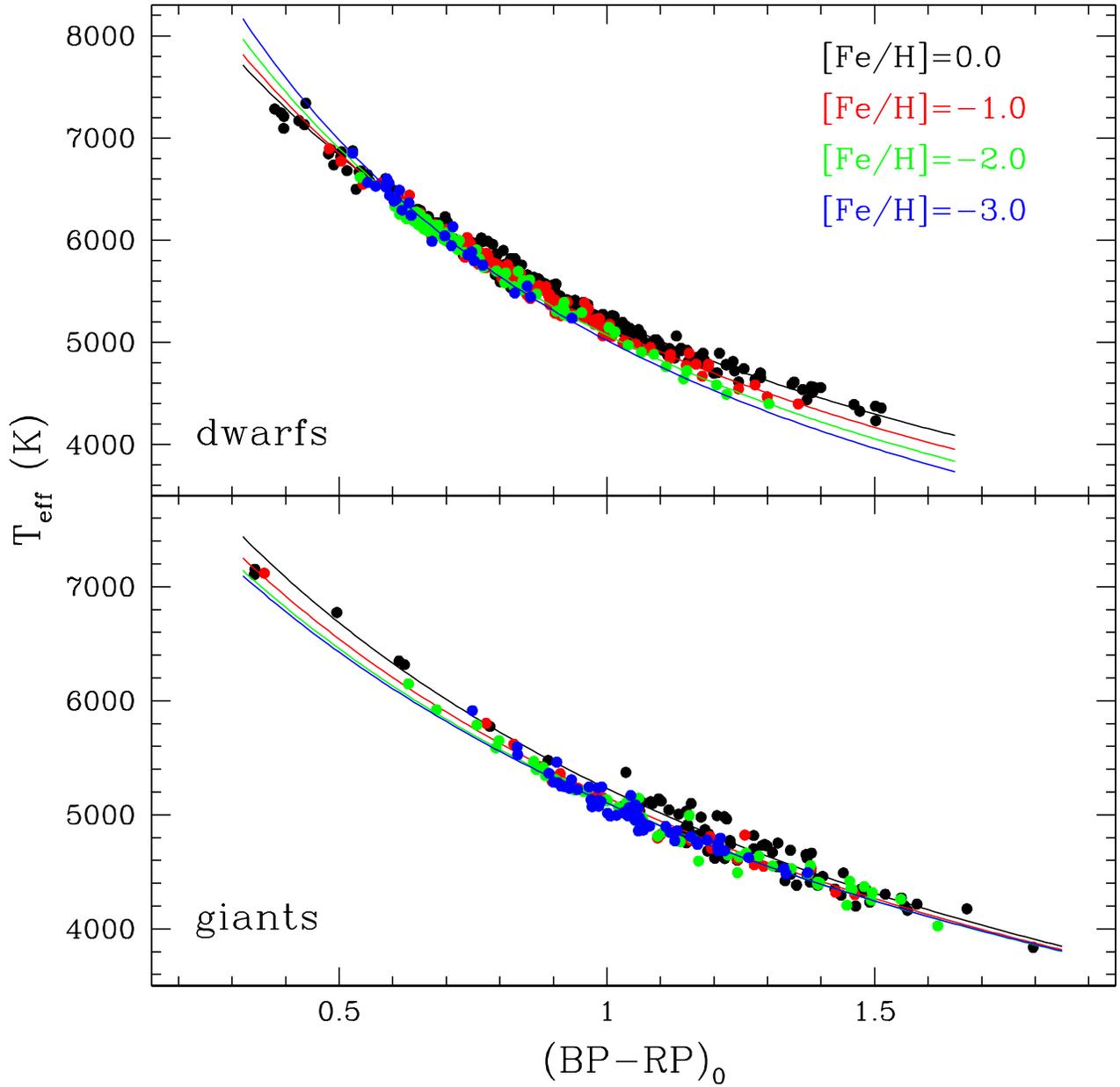}
\caption{Behaviour of \teff\ derived
from IRFM by \citet{ghb09} as a function of the 
\bprp\ colour, for dwarf and giant stars (upper and lower panels, respectively). The stars are grouped according to their metallicity: [Fe/H]$<$--2.5 dex 
(blue points),--2.5$<$[Fe/H]$,$--1.5 (green points), --1.5$<$[Fe/H]$,$--0.5 (red points), 
[Fe/H]$>$--0.5 dex (black points). The solid lines are the theoretical colour-\teff relation calculated with 
[Fe/H]==--3.0 dex (blue line), --2.0 dex (red line), --1.0 dex (red line), +0.0 dex (black line).}
\label{bprp}
\end{figure}

\begin{figure}
\includegraphics[width=\columnwidth]{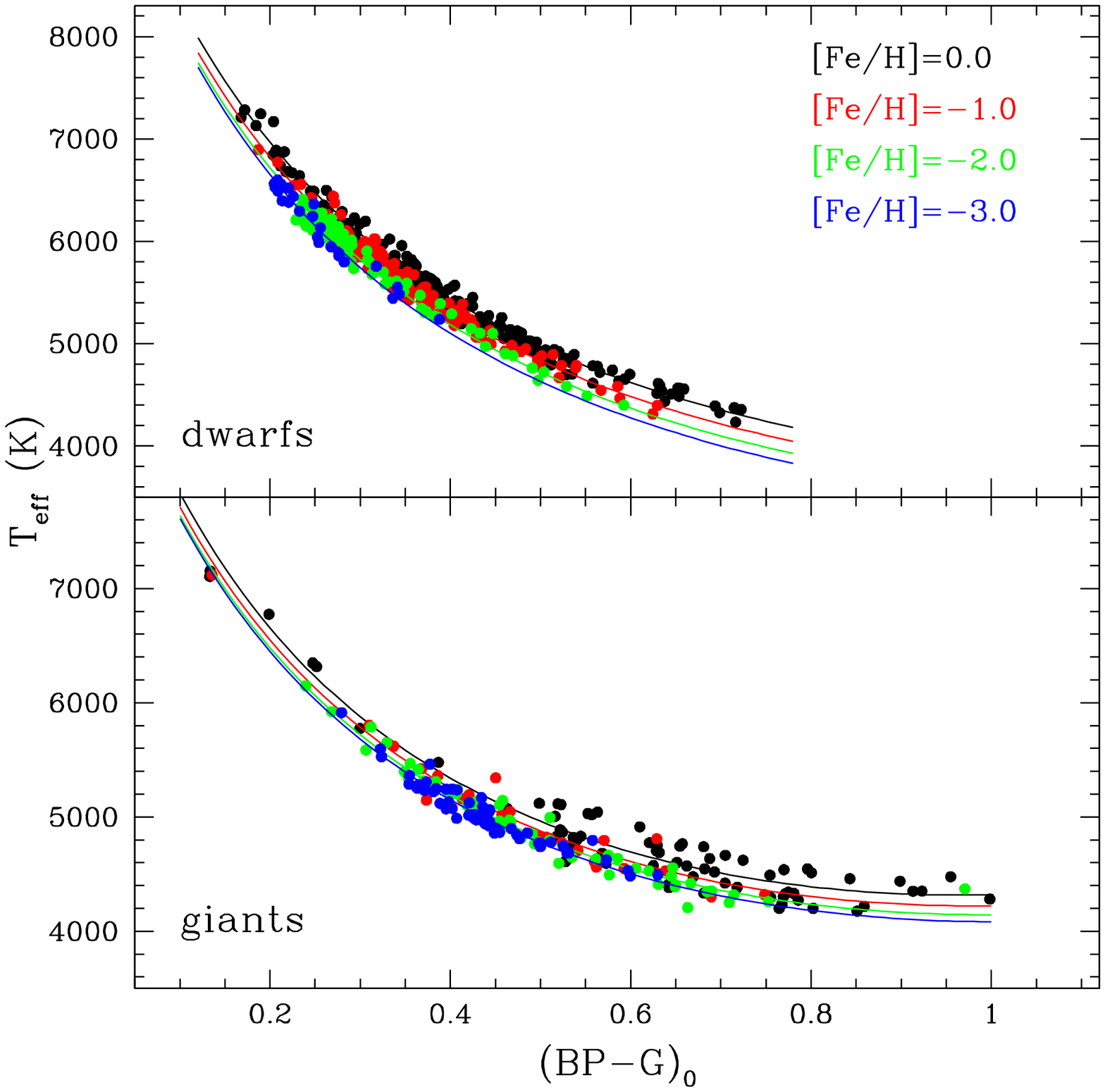}
\caption{Same of Fig.~\ref{bprp} but for the  
\bpg\ colour.}
\label{bpg}
\end{figure}

\begin{figure}
\includegraphics[width=\columnwidth]{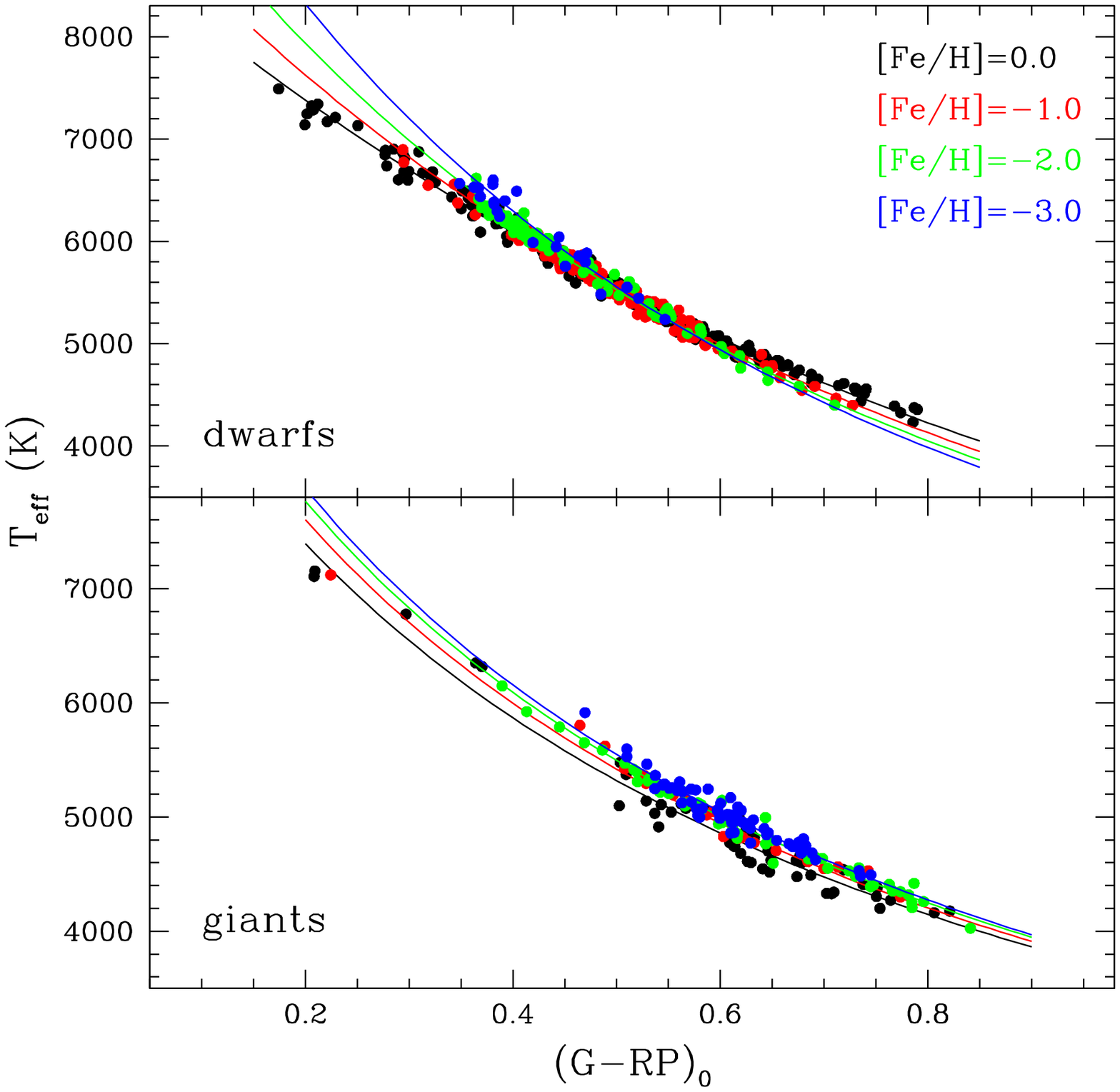}
\caption{Same of Fig.~\ref{bprp} but for the  
\grp\ colour.}
\label{grp}
\end{figure}

\begin{figure}
\includegraphics[width=\columnwidth]{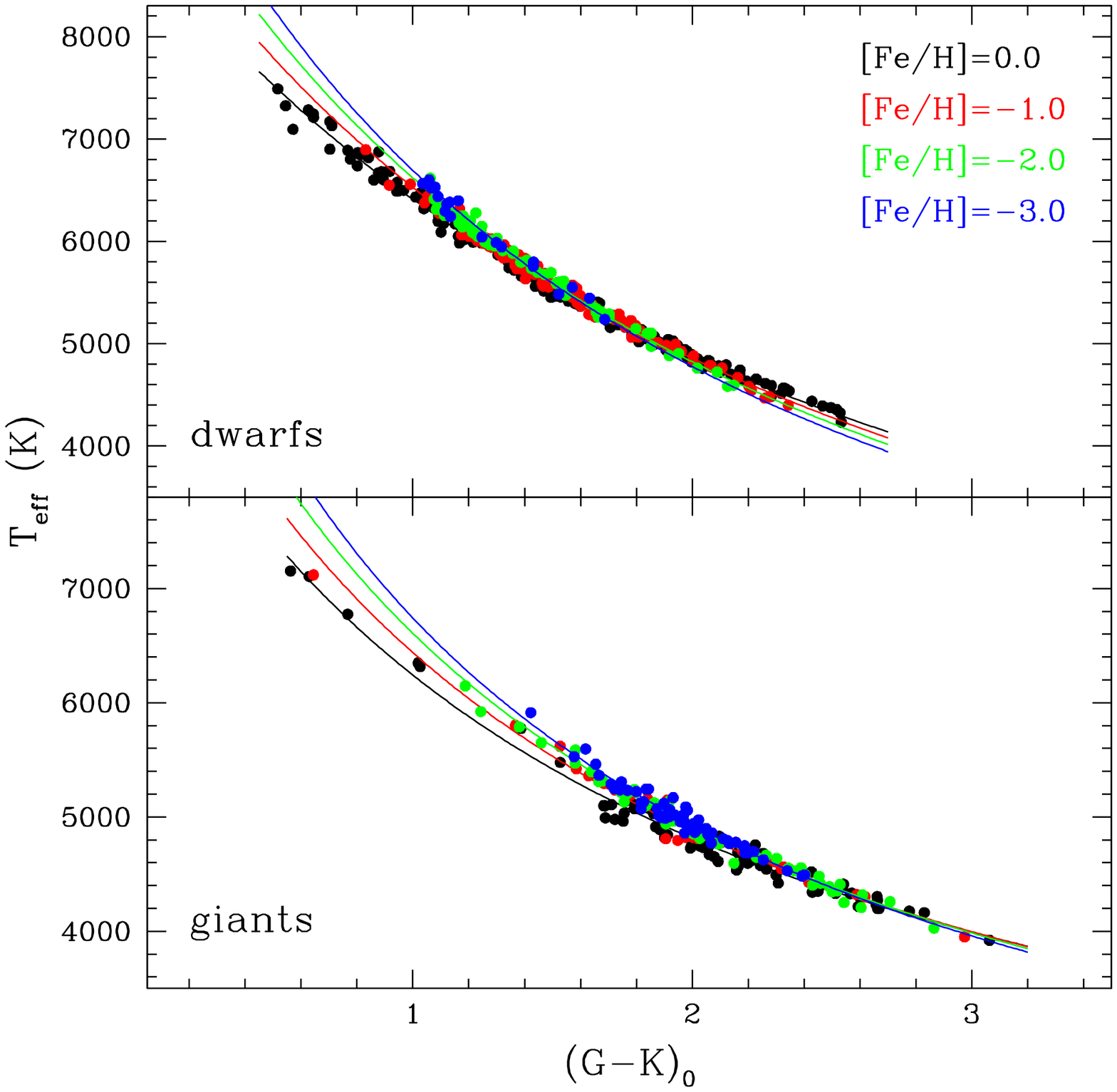}
\caption{Same of Fig.~\ref{bprp} but for the  
\gk\ colour.}
\label{gk}
\end{figure}

\begin{figure}
\includegraphics[width=\columnwidth]{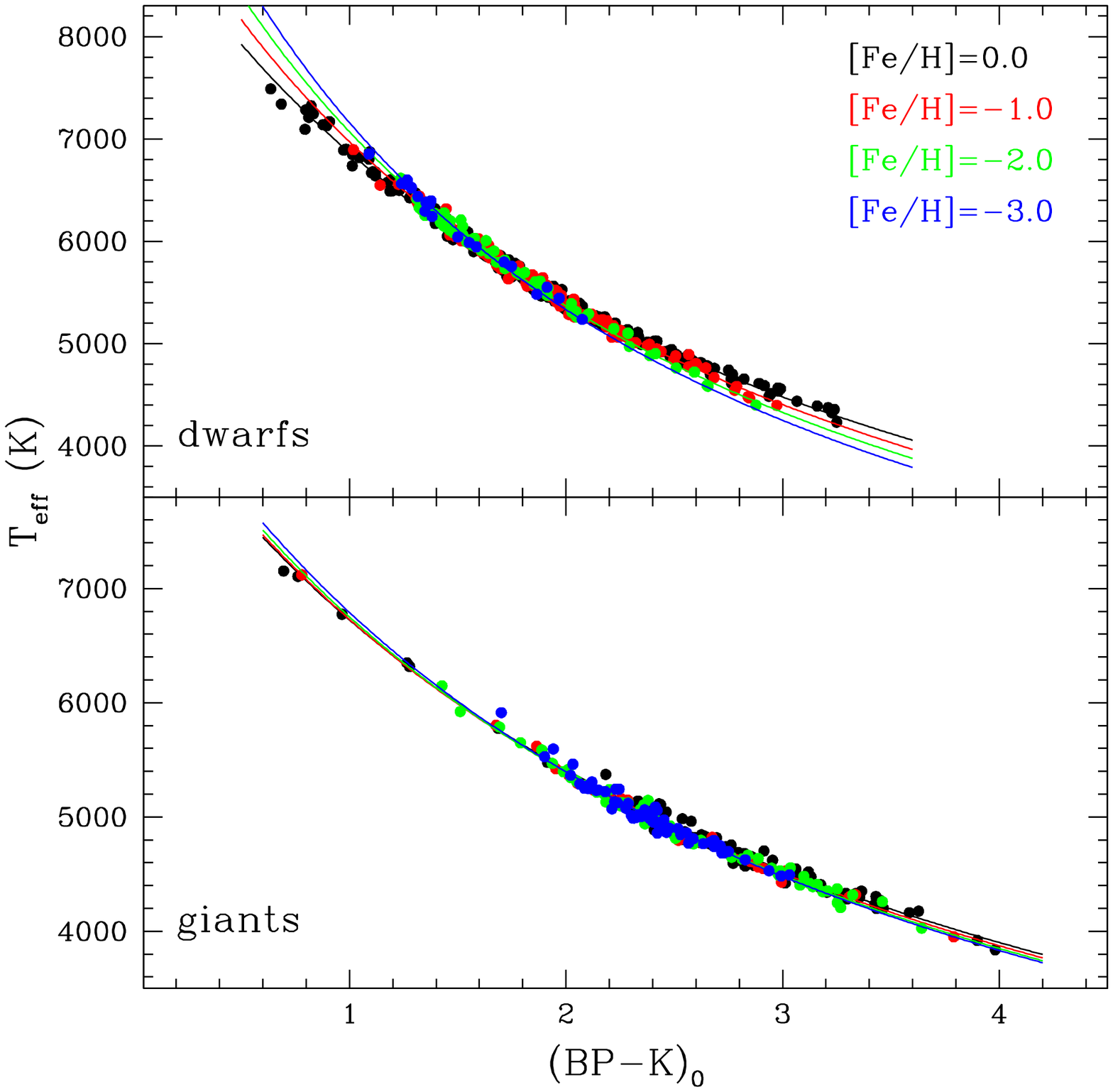}
\caption{Same of Fig.~\ref{bprp} but for the  
\bpk\ colour.}
\label{bpk}
\end{figure}

\begin{figure}
\includegraphics[width=\columnwidth]{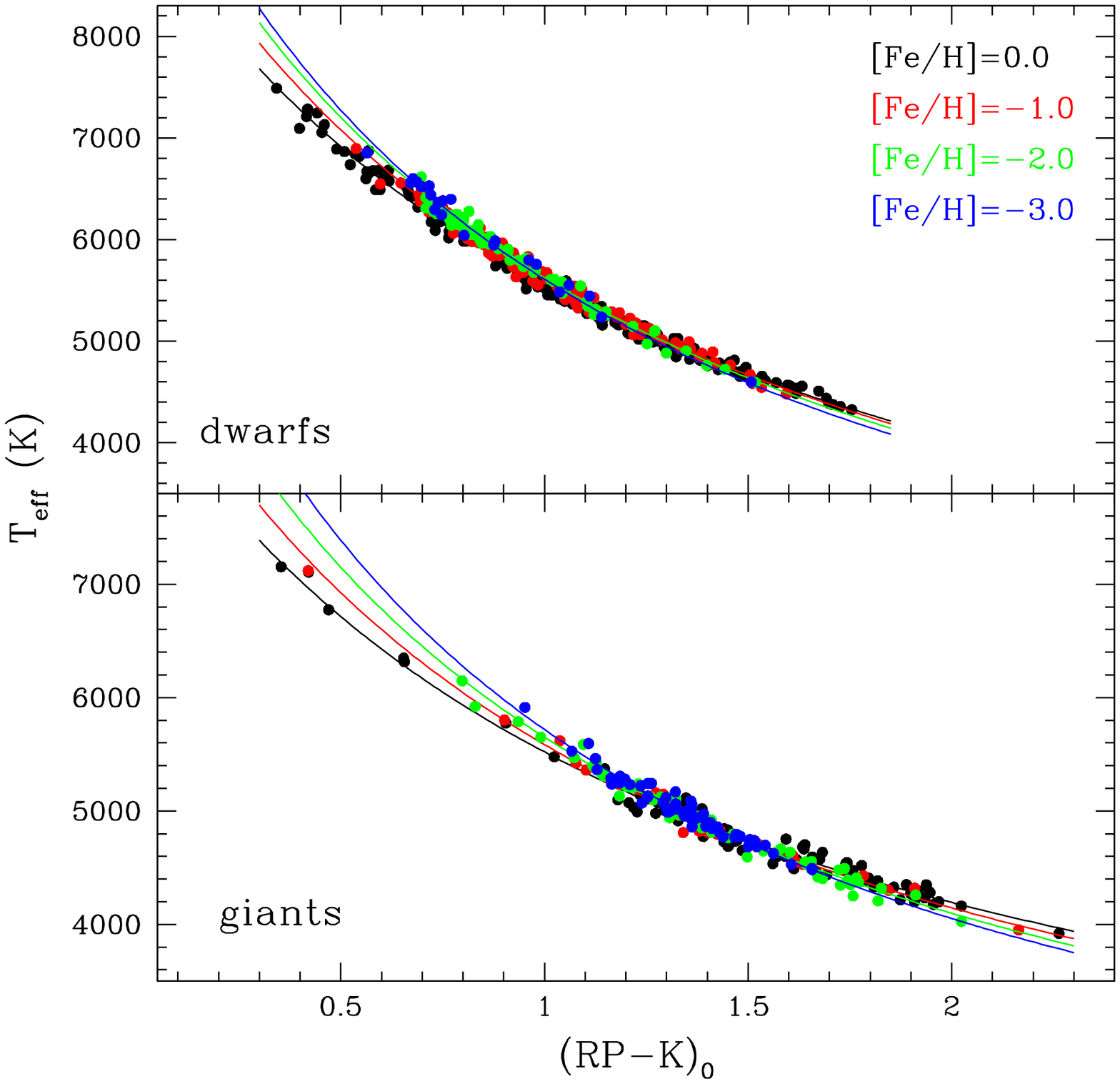}
\caption{Same of Fig.~\ref{bprp} but for the  
\rpk\ colour.}
\label{rpk}
\end{figure}

\begin{figure}
\includegraphics[width=\columnwidth]{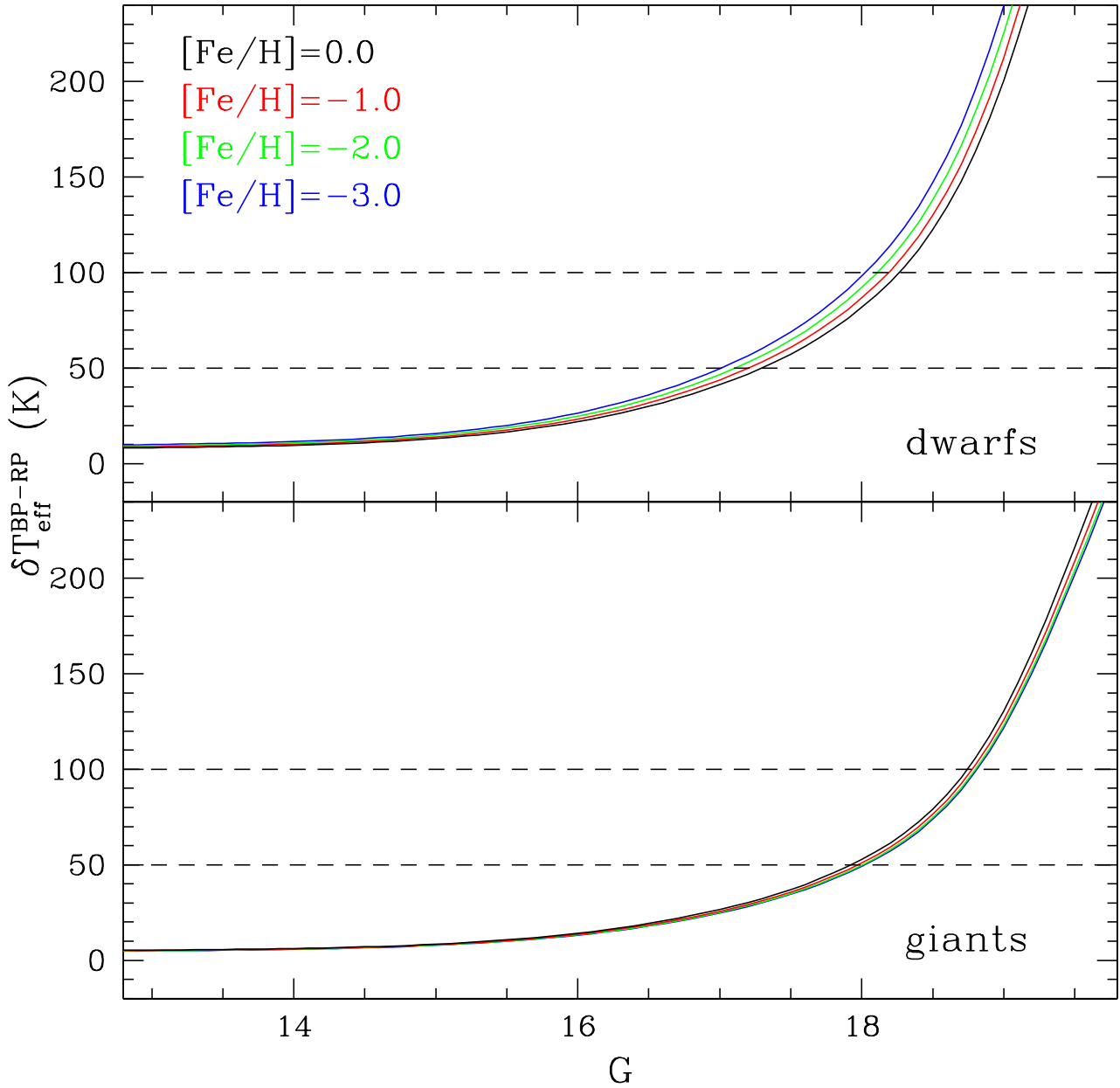}
\caption{Behaviour of the expected error in \teff\ derived from \bprp\ colour as a function of the G-band magnitude 
(upper and lower panels show results for dwarf and giant stars, respectively, adopting the same colour codes used in Fig.~\ref{bprp}). The 
horizontal dashed lines marks +50 K and +100K.}
\label{errteff}
\end{figure}

\end{document}